\documentclass[twocolumn,showpacs,preprintnumbers]{revtex4}
\usepackage{amsmath,amssymb}
\usepackage{graphicx}
\usepackage{dcolumn}
\usepackage{bm}

\begin{document}

\title{Saturated absorption spectroscopy: elimination of crossover resonances by use of a  nanocell}

\author{A. Sargsyan, D. Sarkisyan, A. Papoyan}
\affiliation{Institute for Physical Research, National Academy of Sciences of Armenia, Ashtarak, 0203 Armenia}
\author{Y. Pashayan-Leroy}
\affiliation{Laboratoire de Physique de l'Universit\'{e} de Bourgogne, UMR-CNRS 5027, 21078
Dijon Cedex, France}

\author{P. Moroshkin, A. Weis}

\affiliation{D\'{e}partment de Physique, Universit\'{e} de Fribourg, Chemin du Mus\'{e}e 3, 1700 Fribourg, Switzerland}


\author{A. Khanbekyan, E. Mariotti, and L. Moi}
\affiliation{Department of Physics, University of Siena, Via Roma 56, 53100 Siena, Italy}

\begin{abstract}It is demonstrated that velocity selective optical pumping/saturation resonances of reduced absorption in a Rb vapor nanocell with thickness \textit{L=} $\lambda $, 2$\lambda $, and 3$\lambda $ (resonant wavelength $\lambda $ = 780 nm) allow the complete elimination of crossover (CO) resonances. We observe well pronounced resonances corresponding to the F$_{g}=3$ $\rightarrow $ F$_{e}=2,3,4$ hyperfine transitions of the $^{85}$Rb D$_{2}$ line with linewidths close to the natural width. A small CO resonance located midway between F$_{g}=3$ $\rightarrow $ F$_{e}=3$ and F$_{g}=3$ $\rightarrow$ F$_{e}=4$ transitions appears only for \textit{L} = 4$\lambda $. The D$_{2}$ line ($\lambda $ = 852 nm) in a Cs nanocell exhibits a similar behavior. From the amplitude ratio of the CO and VSOP resonances it is possible to determine the thickness of the column of alkali vapor in the range of 1 - 1000 $\mu $m. The absence of CO resonances for nanocells with $L \sim \lambda $ is attractive for frequency reference application and for studying transitions between Zeeman sublevels in external magnetic fields.
\end{abstract}

\maketitle





\

\maketitle 

\section{INTRODUCTION}

\textbf{}

\textbf{    }Saturated absorption (SA) spectroscopy is widely used in the realization of frequency references for atomic transitions [1-3]. In this technique the laser beam is split into a weak probe field and a strong pump field, which are sent to the interaction cell as counter-propagating overlapping beams. Because of opposite Doppler shifts, only the atoms moving perpendicular to the radiation propagation direction resonantly interact with both laser beams. For these atoms, the pump beam saturates the transition, and the absorption spectrum of the probe shows a Doppler-free dip, the so called velocity selective optical pumping/saturation (VSOP) resonance located at the line center. With properly chosen pump and probe beam intensities, careful adjustment of the geometry, and elimination of stray magnetic fields, the linewidth of the resonance (to which we refer to as "VSOP" resonance) may be as narrow as the natural width of the transition. 

The situation is more complex when two or more atomic (hyperfine) transitions overlap within the Doppler profile, which is the case for the majority of real atomic lines. The presence of multiple unresolved hyperfine transitions results in the formation of so called crossover (CO) resonances. These spurious resonances appear when the laser frequency is tuned midway between two transitions, so that for group of atoms moving with a certain longitudinal velocity the pump beam drives one transition, while the probe drives the other one linked to the former by the same ground state. As a result of the partial depletion of the ground state by the pump beam, the probe "sees" a reduced absorption. Thus, the VSOP process is also responsible for the occurrence of CO resonances. The CO resonances may seriously complicate the applicability of the SA spectral reference technique, notably in high-resolution spectroscopy of hyperfine levels with a small (10 $\div$ 30 MHz) frequency spacing. In that case the CO resonances mask the real atomic VSOP resonances and thus hinder of even impede the identification of the relevant spectral features. We note that, as a rule, the amplitude of the CO resonances are larger than the one of the actual VSOP resonances.

There are several techniques that allow the elimination of crossover resonances in atomic vapors. In [4] two \textit{co-propagating} laser beams (a pump laser with fixed frequency and a probe laser with tunable frequency) were used rather than \textit{two counter-propagating} beams. In [5] it was shown that the use of a thin (20 $ \mu $m-long) cell in conventional SA setup allows a reduction of the amplitude of CO resonance. Note, that the optical selective reflection (SR) spectroscopy technique [6,7] also allows one to eliminate CO resonances, but for a correct determination of an atomic transition position by this technique the spectra must undergo further non-trivial processing.

 The aim of this work is to show that for very small (wavelength-scale) thicknesses \textit{L} of the atomic vapor column it is possible to completely eliminate the CO resonances, both in single-beam transmission spectroscopy and in the SA configuration.

\section{EXPERIMENTAL SETUP }

The scheme of the experimental setup is presented in Fig.1. The beam from a single-frequency tunable diode laser (1 mm diameter, power $P_{L}$ = 30 mW, central wavelength $\lambda $ = 780 nm, linewidth $\gamma_{L} \sim $ 5 MHz) irradiated a thin cell \textit{1} under an incidence angle close to the normal. 
\begin{figure}[htbp]
 \includegraphics[width=0.5\textwidth]{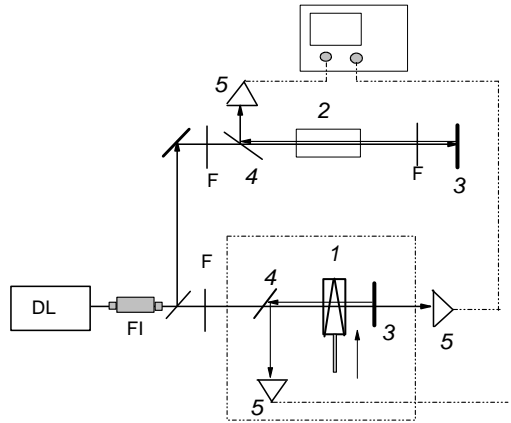}
 \caption{Experimental setup. \textit{DL}- tunable diode laser, $\lambda $ = 780 nm (852 nm); 
\textit{FI}- Faraday isolator; \textit{1-} Rb (Cs) cell with nanometric, micrometric or ordinary thickness
 (see the text); \textit{F}- neutral density filters; \textit{2}- Rb (Cs) cell 
of ordinary length; \textit{3-  }mirror;\textit{ 4-} glass plate; \textit{5}- photodiodes.}
 \end{figure}
A Faraday isolator (\textit{FI}) was inserted to prevent optical feedback to the laser. In the experiment we have used different cells \textit{1} with various thicknesses \textit{L} in the ranges: \textit{L }= $\lambda $ $\div$ 6$\lambda $ (nanometric, vertical wedged-shaped vapor column) [8-13]; 9 $\mu$m, 60 $\mu$m, 700 $\mu$m [14]; and 2 mm, 4 mm, 60 mm-long. The cell was inserted into an oven with two apertures for passing the radiation. The Rb density determining temperature of the cell's side-arm was adjusted in the range of 40 $\div$ 115 $^{o}$C, depending on the cell thickness. To prevent atomic vapour condensation on the cell windows their temperature was kept 20 to 30 degrees higher that the sidearm tempearture. The laser intensity was controlled by neutral density filters \textit{F} and its  frequency scanned over $\sim $ 10 GHz, covering the hyperfine spectral components of the the $^{85}$Rb D$_{2}$ lines F$_{g}=3$ $\rightarrow $ F$_{e}=2,3,4$, where the subscripts \textit{g}, \textit{e} refer to the lower and upper levels, respectively). A mirror \textit{3 }and\textit{ }a glass plate \textit{4 }(which are shown in Fig.1 inside the dotted rectangle) were used to form the SA spectra. In order to vary thicknesses L of the vapour column in the nanocell in the range from \textit{L }= (1 -- 6) $\lambda $ the oven was translated along the vertical direction as indicated by the arrow in Fig.1. A part of the laser beam was sent to a normal vapor cell with a thickness of 60 mm to obtain a reference SA spectrum. The upper set of filters \textit{F} was used to properly attenuate the laser beam in order to get the reference SA spectra with a linewidth close to the natural width. The spectra were registered in a four-channel digital storage oscilloscope (Tektronix, TDS 2014B).

\section{RESULTS AND DISCUSSION}

\textbf{ }

It was shown earlier [9-13] that the ratio \textit{L}/$\lambda $, where \textit{L }is the thickness of the atomic vapor column and $\lambda $ the laser wavelength resonant with the atomic transition, is an important parameter that determines the widths, shapes and amplitudes of the absorption resonances in a nanocell. In particular, it was shown that the spectral width of resonant absorption is minimal for \textit{L }= (2\textit{n }+1) $\lambda $/2 (where \textit{n} is an integer), an effect which was called "Dicke-type Coherent Narrowing Effect" (DCNE). It was also shown that for \textit{L }= \textit{n}$\lambda $ the spectral width of the resonant absorption reaches a maximal value close to the Doppler width (about several hundreds of MHz), an effect called collapse of DCNE [9,10]. In [11] the DCNE effect and the collapse of DCNE were investigated up to the thicknesses \textit{L }= 7$\lambda $/2 for the D$_{2}$ line of $^{85}$Rb at $\lambda $ = 780 nm. Here we consider the case where \textit{L }= \textit{n}$\lambda $, since as it was shown in [10,11] that the VSOP resonances  appear under  this condition. 
\begin{figure}[htbp]
 
 \includegraphics[width=0.5\textwidth]{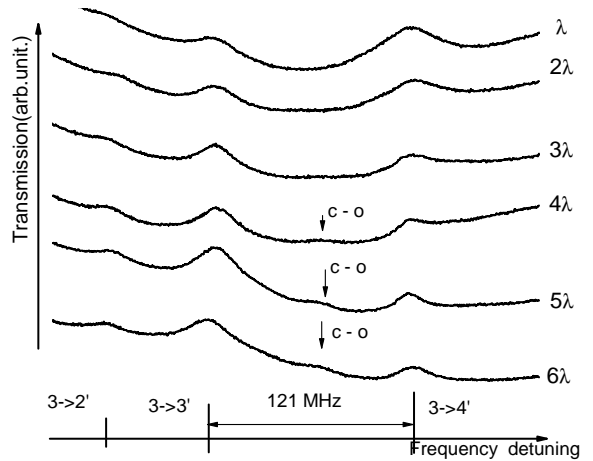}

 \caption{Transmission spectra of the nanocell for the transitions F$_{g}=3 \rightarrow  F_{e}=2,3,4$ of $^{85}$Rb D$_{2}$ line, for \textit{L} varying in the range $\lambda $ $\div$ 6 $\lambda $. The effective laser intensity was $\sim $ 10\textit{ }mW/cm$^{2}$. The temperature of the cell's side-arm is 115 $^{o}$C. The CO resonances are marked by arrows.}
 \end{figure}
In Fig.2 the transmission spectra are presented for \textit{L} varying in the range from \textit{L }= $\lambda $ up to 6$\lambda $ with steps of $\lambda $ (for these recordings the mirror \textit{3} and the glass plate \textit{4} were removed). The effective laser intensity (EI) is determined by multiplying the measured intensity by the coefficient $[\gamma_{N}/(\gamma_{N} +\gamma_{L}]$, where $\gamma_{N} $ is the spontaneous decay rate of the Rb D$_{2}$  line, $\gamma_{L} \sim $5 MHz,\textit{ }and EI is\textit{ }10\textit{ }mW/cm$^{2}$. When \textit{L }is\textit{ }varied from \textit{L} = $\lambda $ to 3$\lambda $ it can be seen that only the VSOP resonances are detected. These peaks of decreased absorption are located exactly at the the atomic transition frequencies [5,10-14], and arise because the atom in the ground level F$_{g}=3$ absorbs a laser photon populating the excited level, followed by spontaneous decay to the ground level F$_{g}=2$ or F$_{g}=3$, an effect well known as optical pumping (OP) [1-3,5,15] (peculiarities of OP process in sub-millimeter thin cell are presented in [16,17]). As a result, a fraction of the atoms populates the F$_{g}=2$ level, and the number of atoms absorbing from the F$_{g}=3$ level is reduced. As a consequence absorption from this level decreases. The efficiency of OP is determined by the expression\textit{}      

\begin{equation} \label{GrindEQ__1_} \eta \sim \frac{\Omega ^{2} \gamma_{N} t}{\left(\Delta +\vec{k}\vec{v}\right)^{2} +\Gamma ^{2} }  \end{equation} 

where \textit{t }is the average interaction time of the atom with the radiation field, \textit{v} the atomic velocity, $\Delta $  the detuning, $\Gamma $ the sum of homogeneous and inhomogeneous broadenings, and \textit{k = 2$\pi $/}$\lambda $ [15]\textit{.} Eq. \eqref{GrindEQ__1_} shows how the optical pumping efficiency grows with the interaction time \textit{t}. For atoms flying perpendicularly to the laser beam the interaction time is $t_{D}= D/v$, where \textit{D }is the laser beam diameter, while atoms flying along the laser beam have an interaction time of $t_{L} = L/v$. Since D $\sim $ 1 mm, and \textit{L }= 780 nm(or 852 nm in the case of Cs), $t_{D}$ exceeds $t_{L}$ by three orders of magnitude. For atoms flying perpendicular to the laser beam \textbf{k·v }= 0, and the efficiency in Eq. \eqref{GrindEQ__1_} becomes maximal for $\Delta $ = 0. For this reason the VSOP peak is centered exactly at the atomic transition
\begin{figure}[htbp]
 
 \includegraphics[width=0.5\textwidth]{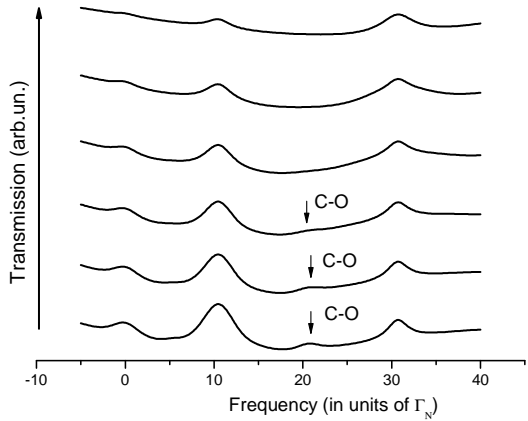}

 \caption{Calculated spectra for the conditions of Fig.2. Rabi frequency $\Omega $ = 0.4 $\gamma_{N}$; $\gamma_{L} = 5$ MHz; $\gamma_{N}  = 6$ MHz.}
 \end{figure}
 frequency [5,10-14]. Note, that there is a reflected beam from the inner surface of the nanocell which propagates backwards with respect to the main beam, and this beam may cause the formation of CO resonance. However, the interaction time $t_{L}$  for atoms with a longitudinal velocity $v_{z}$ = 2\textit{$\pi $}$\varepsilon $/\textit{k } (where $\varepsilon $ is half of the frequency separation of corresponding excited levels) along the laser beam is rather small for  \textit{L=} $\lambda $, 2$\lambda $, and 3$\lambda $ to provide an efficient optical pumping which is needed for CO resonansce formation (in our case 2$\varepsilon $ is 121 MHz and 63 MHz, respectively).

However, as can be seen from Fig.2, a small CO resonance (marked by arrows) located midway between the F$_{g}=3 \rightarrow F_{e}=3$ and F$_{g}=3 \rightarrow F_{e}=4$ transitions appears in the transmission spectra when the thickness \textit{L} $=$ 4$\lambda $. One can also see a different behavior of VSOP peaks at different transitions. It is interesting to note that when \textit{L }= $\lambda $ the amplitude ratios of the VSOP peaks is close to the ratio of the atomic probabilities of the corresponding transitions F$_{g}=3 \rightarrow F_{e}=2,3,4$, i.e. \textit{A}(3-4$'$)/\textit{A}(3-3$'$) $\sim$ 2, \textit{A}(3-3$'$)/\textit{A}(3-2$'$) $\sim$ 3 [18]. As the thickness increases, the amplitude of the VSOP peak corresponding to the F$_{g}=3 \rightarrow F_{e}=4$ cycling transition decreases, while the amplitudes of the VSOP peaks for the non-cycling transitions F$_{g}=3 \rightarrow F_{e}=2,3$ increase. In particular, for \textit{L }= 6$\lambda $ the ratio \textit{A}(3-4$'$)/\textit{A}(3-3$'$) $\sim$ 0.5. Conversely, the ratio of the\textit{ }non-cycling transitions is practically independent of the thickness, i.e. \textit{A}(3-3$'$)/\textit{A}(3-2$'$) $\sim$ 3. Fig.3 presents the theoretical spectra for the same group of transitions, with the Rabi frequency $\Omega = 0.4 \gamma_{N}$, $\gamma_{L} = 5$ MHz, and $\gamma_{N}= 6$ MHz. One sees that the theoretical model presented in [19] correctly describes the peculiarities of the VSOP of the cycling and non-cycling transitions as well as the behavior of the CO resonance as a function of the thickness \textit{L}. One can conclude from Figs.2 and 3 that the optimum condition for the formation of an atomic reference spectrum (e.g., to determine the frequency position of a weak atomic transition as F$_{g}=3 \rightarrow F_{e}=2)$ is obtained for \textit{L }= 3$\lambda$, in which case narrow and large VSOP peaks are observed, along with the absence of CO resonances.

The second set of measurements was done for the same group of transitions employing SA configuration (mirror \textit{3} and glass plate \textit{4} are mounted as shown in Fig.1). Thin cells with thicknesses \textit{L }= $\lambda $ (780 nm), 5$\lambda $ (3.9 $\mu$m), 9 $\mu$m, 60 $\mu$m, 700 $\mu$m and cells with an length of 2 mm, 4 mm and 60 mm have been used.
\begin{figure}[htbp]
 
 \includegraphics[width=0.5\textwidth]{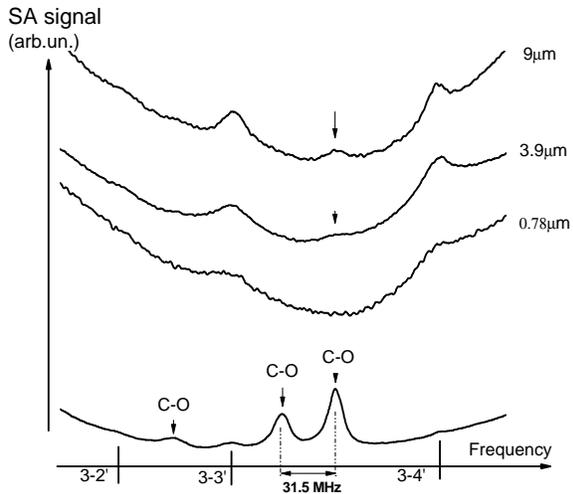}

 \caption{Saturated absorption spectra for Rb cells with \textit{L }= 780 nm, 3.9 $\mu$m, 9 $\mu$m. The lower curve is the SA spectrum of an ordinary 60 mm long cell. The CO resonances are marked by arrows.}
 \end{figure}
In Fig.4 the SA spectra are presented for \textit{L }= $\lambda $ (780 nm), 5$\lambda $ (3.9 $\mu$m), 9 $\mu$m. The lower curve is the SA spectrum recorded in a 60 mm long cell serving as a reference. As before the CO resonance is completely absent in the case of \textit{L }= $\lambda $, while it appears when \textit{L }= 5 $\lambda $ (3.9 $\mu$m) and the amplitude of the CO resonance increases as \textit{L }increases from 3.9 $\mu$m to 9 $\mu$m.
\begin{figure}[htbp]
 
 \includegraphics[width=0.5\textwidth]{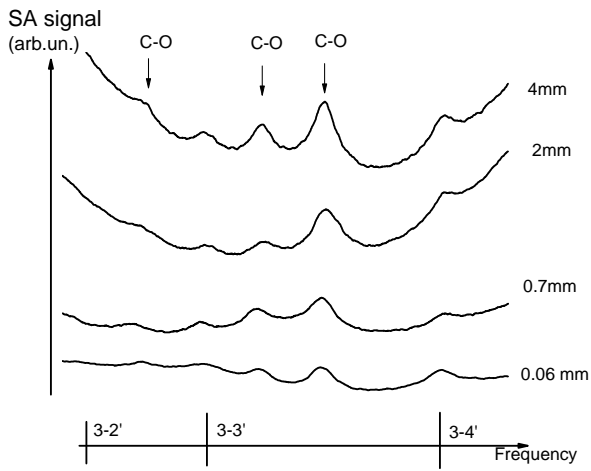}

 \caption{Saturated absorption spectra for Rb cells with \textit{L }= 60 $\mu$m, 0.7 mm, 2 mm, and 4 mm.}
 \end{figure}
Fig.5 shows the SA spectra for \textit{L }= 60 $\mu$m, 0.7 mm, 2 mm, and 4 mm. The VSOP resonance of the F$_{g}=3$ $\rightarrow $ F$_{e}=4$ transition is sensitive to the polarizations of the pump and probe beams [20]. For a quantitative description it is therefore more convenient to use the ratio of the amplitude of the CO resonance appearing when the laser is tuned midway between F$_{g}=3 \rightarrow  F_{e}=3$ and F$_{g}=3 \rightarrow F_{e}=4$ atomic transitions (this is the largest CO resonance in the spectrum), and the amplitude of the VSOP resonance of the  F$_{g}=3 \rightarrow F_{e}=3$  transition. Fig.6 shows the ratio of \textit{A}(CO)/\textit{A}(VSOP 3-3$'$) as a function of the cell thickness. The monotonic smooth increase of this dependence permits to infer the cell thickness from a measurement of the amplitude ratio. Note, that in the case of thinner cells (\textit{L} $\rightarrow \lambda $) the interferometric method presented in [9] allows the determination of the thickness \textit{L} with high accuracy $\sim $ 15 nm). It should be noted that the dependence shown in Fig.6 is slightly affected by the pump and probe beam intensities, and is correct for the $^{85}$Rb, F$_{g}=3 \rightarrow F_{e}= 2,3,4$ atomic transitions.

There are several reasons which influence the CO resonance amplitude, in particular the frequency separation between the upper levels (2$\varepsilon $), the probabilities of the atomic transitions involved in the CO resonance formation,  the thermal atomic velocity of the alkali atom [2,3]. A frequency-tunable distributed feedback diode laser with $\lambda $ = 852 nm (linewidth 5 MHz) and EI $\sim $10 mW/cm$^{2}$  has been used to obtain  the transmission spectra (see Fig.7) for a nanocell filled with Cs vapor for L varying in the range of  $\lambda $ to 6 $\lambda $ in steps of $\lambda $ (F$_{g}= 4 \rightarrow  F_{e}=3,4,5$ transitions of D$_{2}$ line).  In this case a small CO resonance appears at $L = 6 \lambda$ when the laser is tuned midway between the F$_{g}=4 \rightarrow F_{e}=4$ and F$_{g}=4 \rightarrow F_{e}=5$ atomic transitions. This is in agreement with the results presented in [5], where the ratio \textit{A}(CO)/\textit{A}(VSOP 4-4$'$)  is $\sim $ 0.4 for the F$_{g}=4 \rightarrow F_{e}=3,4,5$ transitions in Cs and \textit{L} = 100 $\mu$m, while for the F$_{g}=3 \rightarrow F_{e}=2,3,4$  group in $^{85}$Rb the ratio \textit{A}(CO)/\textit{A}(VSOP 3-3$'$) is $\sim $ 2 at an even smaller cell thickness (\textit{L} = 60 $\mu$m) (see Fig.5).
\begin{figure}[htbp]
 
 \includegraphics[width=0.5\textwidth]{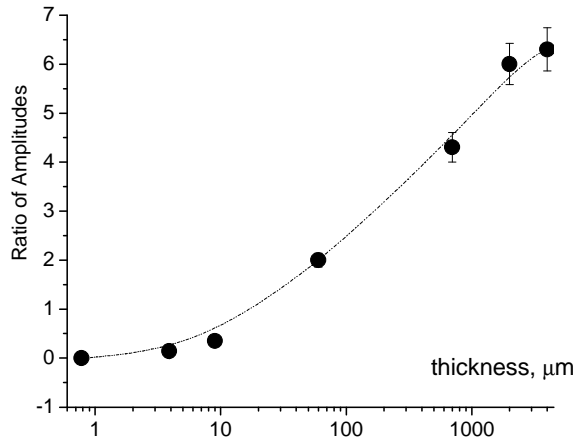}

 \caption{Amplitude ratios \textit{A}(CO)/\textit{A}(VSOP 3-3$'$) as a function of the cell thickness. The dotted line is shown to guide the eye.}
 \end{figure}
\begin{figure}[htbp]
 
 \includegraphics[width=0.5\textwidth]{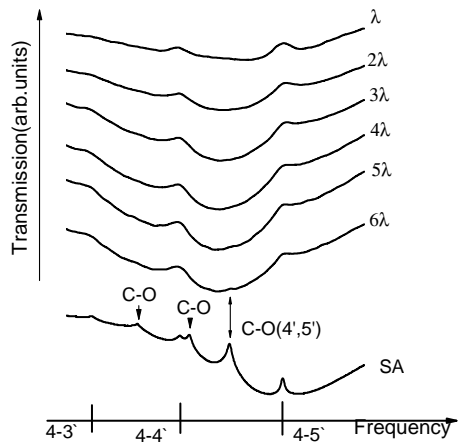}

 \caption{Transmission spectra of the Cs nanocell for the transitions F$_{g}=4 \rightarrow  F_{e}=3,4,5$ of $^{133}$Cs D$_{2}$ line, for \textit{L} varying in the range  $\lambda $ $\div$  6 $\lambda $. The effective laser intensity was $ \sim  $ 10\textit{ }mW/cm$^{2}$.  The temperature of the side-arm was 110 $^{o}$C. The lower curve is the SA spectrum in an ordinary cell. The CO resonances are marked by arrows.}
 \end{figure}
The difference of the SA spectra for different atomic transitions is well seen if we compare spectra obtained in similar conditions (i.e., cell length, temperature, laser intensity, etc.), as is the case for the lower traces of Figs. 4,7 ($^{85}$Rb D$_{2}$ line, F$_{g}=3 \rightarrow F_{e}=2,3,4$ in Fig.4, and Cs D$_{2}$ line, F$_{g}=4 \rightarrow F_{e}= 3,4,5$ in Fig.7). For $^{85}$Rb the ratio \textit{A}(CO)/\textit{A}(VSOP 3-3$'$) is $\sim $ 10, while for Cs we find \textit{A}(CO)/\textit{A}(VSOP 4-4$'$) $\sim $ 2. Similar ratios can be also seen in [2,3]. From this behavior one can expect that for the case of the Cs F$_{g}= 4 \rightarrow F_{e}=3, 4,5$ transitions the increase of the CO resonance amplitude with the thickness should be somewhat weaker. 
\begin{figure}[htbp]
 
 \includegraphics[width=0.5\textwidth]{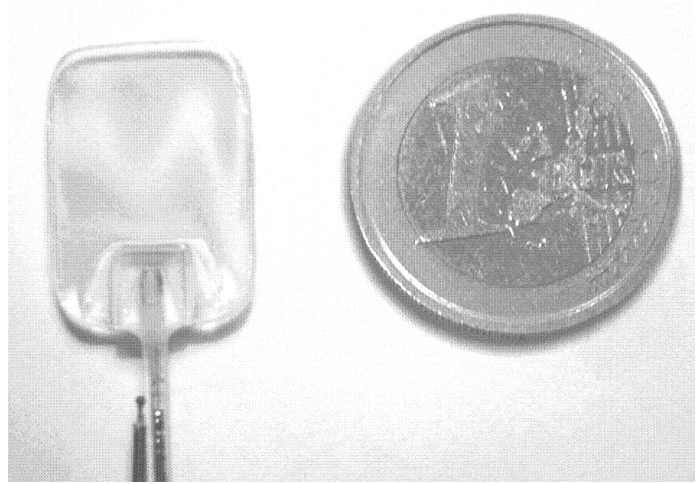}

 \caption{Photograph of a nanocell filled with Na. A 1 Euro coin is shown for the scaling.}
 \end{figure}
 It will be interesting to study the dependence of amplitudes ratio \textit{A}(CO)/\textit{A}(VSOP) on the atomic D$_{1}$ line of Na, since the value of 2$\varepsilon $ in this case is 190 MHz,  i.e., is close to that of $^{85}$Rb, while the atomic velocity for Na is approximately three times larger. For this purpose we have recently prepared a nanocell filled with Na metal (Fig.8). The thickness \textit{L} of the gap between the windows has a wedge in the vertical direction, varying from 100 nm to 550 nm at room temperature. At the windows' operating temperature of $\sim $ 200 $^o$C the maximum thickness \textit{L} slightly increases, reaching 800 nm. Thus, there are two important regions of the thickness \textit{L} = $\lambda $/2 and \textit{L} = $\lambda $ (for Na, $\lambda$ = 590 nm), where sub-Doppler absorption and fluorescence spectroscopy can be realized. This experiment is in progress at Siena University.

\section{CONCLUSION}

We have demonstrated that the use of a nanocell with a thickness \textit{L=} $\lambda $, 2 $\lambda $ and 3$\lambda $ allows one to completely eliminate the CO resonance, while VSOP resonances located on atomic transitions are well pronounced. The advantage of the use of nanocell with thickness $L = \lambda$ as a frequency reference for an atomic transition is that the ratio of the amplitudes for the VSOP peaks is close to the ratio of the atomic probabilities of the corresponding transitions. If, however, the center frequency of a weak atomic transition has to be determined, the use of a nanocell with a thickness $L = 3 \lambda$ is more appropriate. If the thickness \textit{L} = 4$\lambda $ for $^{85}$Rb (\textit{L} $=$ 6 $\lambda $ for $^{133}$Cs), a small CO resonance appears in the transmission spectra of $^{85}$Rb, located midway between the F$_{g}=3 \rightarrow F_{e}=3$ and F$_{g}=3 \rightarrow F_{e}=4$  transitions (F$_{g}=4 \rightarrow F_{e}=4$ and F$_{g}=4 \rightarrow  F_{e}=5$ in the case of Cs).

For spectral reference applications the use of a single-beam transmission spectrum of a nanocell has obvious advantages compared to the SA geometry, which requires counter-propagating beams in the nanocell.

From the ratio of the amplitudes of the CO and VSOP resonances, it is possible to determine the thickness of the alkali vapor column in the range from 1 to  1000 $\mu $m. 

 It is important to note that the absence of CO resonances in a nanocell with \textit{L} $\sim $ $\lambda $ allows one to study the behavior of atomic transitions between the Zeeman sublevels in external magnetic field [21], since the short time of flight ensures that the CO resonances are absent, even when the VSOP resonance is split into several Zeeman components. Conversely, even if CO resonances are absent in SA spectrum of an ordinary cell (this happens when the frequency distance between upper levels is larger than the Doppler width of the atomic transition), the CO resonance may appear again (due to a large time of flight) when the VSOP resonance is split into a several components.

\section{Acknowledgments}

D.S. is grateful to University of Siena for the hospitality. The authors A.S, D.S, A.P, Y.P-L, E.M. and L.M. are thankful for the financial support provided by INTAS South-Caucasus Grant 06-1000017-9001. A.S, D.S, A.P, P.M. and A.W. are thankful for financial support provided by SCOPES Grant IB7320-110684/1.\\

\textbf{REFERENCES}

\textbf{}

\begin{enumerate}
\item \textbf{ }W. Demtröder, \textit{Laser Spectroscopy }(Springer-Verlag, Berlin, 1982).
\item  O. Schmidt, K.-M. Knaak, R. Wynands, and D. Meschede, Appl. Phys. B. \textbf{59},167 (1994) 
\item  D.A. Smith and I.G. Hughes, Am. J. Phys. \textbf{72},\textbf{ }631 (2004).
\item  A. Banerjee and V. Natarajan, Opt. Lett. \textbf{28}, 1912 (2003). 
\item  S. Briaudeau, D. Bloch, and M. Ducloy, Phys. Rev. A \textbf{59}, 3723 (1999). 
\item   R. M$\ddot{u}$ller and  A. Weis, Appl. Phys. B \textbf{66} (3), 323 (1998).
\item  H. Failache,S. Saltiel, M. Fichet , D.Bloch, and M.Ducloy, Phys. Rev. Lett.\textbf{ 83}, 5467 (1999).
\item  D. Sarkisyan, D. Bloch, A. Papoyan, and M. Ducloy, Opt. Commun. \textbf{200}, 201 (2001).
\item  G. Dutier, A. Yarovitski \textit{et al.}, Europhys. Lett. \textbf{63} (1), 35  (2003).
\item  D. Sarkisyan, T. Varzhapetyan \textit{et al.}, Phys. Rev. A \textbf{69}, 065802 (2004).
\item  D. Sarkisyan, T. Varzhapetyan, A. Papoyan, D. Bloch, and M. Ducloy, Proc. SPIE \textbf{6257}, 625701 (2006).
\item  A. Sargsyan, D. Sarkisyan, and A. Papoyan, Phys. Rev. A \textbf{73}, 033803 (2006).
\item  C. Andreeva, S. Cartaleva \textit{et al.}, Phys. Rev. A \textbf{76}, 013837 (2007).
\item  D. Sarkisyan, A. Sargsyan, A. Papoyan, and Y. Pashayan-Leroy, Proc. SPIE \textbf{6604}, 660405 (2007).
\item  G. Nikogosyan, D. Sarkisyan, and Yu. Malakyan, J. Opt. Technol. \textbf{71}, 602 (2004).
\item A.Ch. Izmailov, K. Fukuda, M. Kinoshita, M. Tachikawa, Laser Phys. \textbf{14}(1), 30 (2004).
\item A.Ch. Izmailov,  \textit{Laser Phys. Lett. } \textbf{3}(3), 132  (2006).
\item D. Sarkisyan, T. Becker, A. Papoyan, P. Thoumany, and H. Walther, Appl. Phys. B \textbf{76},  625 (2003).
\item A. Sargsyan, D. Sarkisyan, Y. Pashayan-Leroy, C. Leroy, P. Moroshkin, and A. Weis, http://eprintweb.org/S/archive/physics/0707.0379.
\item C.P. Pearman, C.S. Adams, \textit{et al.}, J. Phys. B: At. Mol. Opt. Phys. \textbf{35}, 5141 (2002).
\item T. Varzhapetyan, H. Hakhumyan, D. Sarkisyan, V. Babushkin, A. Atvars, and M.   Auzinsh, J. Contemp. Phys. \textbf{42}, 223 (2007).
 \end{enumerate}

\end{document}